\documentclass[twocolumn,superscriptaddress,floatfix]{revtex4-2}

\usepackage{verbatim}
\usepackage{hyperref,natbib}
\usepackage{amssymb}
\usepackage{mathrsfs}
\usepackage{amsthm,bbold}
\usepackage{amsmath}
\usepackage{latexsym}
 
\usepackage{subfigure} 
\usepackage{xcolor}
\usepackage{array,float}

\usepackage{optidef} 
 \usepackage{pifont,array}

 \usepackage{colortbl}
 \makeatletter
\def\thickhline{%
  \noalign{\ifnum0=`}\fi\hrule \@height \thickarrayrulewidth \futurelet
   \reserved@a\@xthickhline}
\def\@xthickhline{\ifx\reserved@a\thickhline
               \vskip\doublerulesep
               \vskip-\thickarrayrulewidth
             \fi
      \ifnum0=`{\fi}}
\makeatother

\newlength{\thickarrayrulewidth}
\newcolumntype{P}[1]{>{\centering\arraybackslash}p{#1}}

\usepackage{xspace}

\def\ba{{\bf a}} 
\def\bb{{\bf b}} 
 
\def\bx{{\bf x}}

\def\cC{{\mathcal C}} 

\def\logical{{\mathrm{L}}} 

\def\>{\rangle} 
\def\<{\langle}

\def\lncy{{\texttt{LNCY}}\xspace}
\def\abc{{\texttt{ABC+}}\xspace}
\def\new{{\texttt{8qubit}}\xspace}
\def\rep{{\texttt{REP2}}\xspace}
\def\klm{{\texttt{KLM}}\xspace}
\def\concat{{\texttt{2LNCY}}\xspace}

\def\stab{{\texttt{Stab}}\xspace}
\def\decoder{{\texttt{Dec}}\xspace}
\def\paulirot{{R}}

\def\effgam{\epsilon_{\rm base}}

\DeclareMathOperator{\bin}{bin}

\DeclareMathOperator{\wt}{wt}

\newcommand{\lncycode}{$\mathcal C_{\lncy}$\xspace} 
\newcommand{\repcode}{$\mathcal C_{\rep}$\xspace}
\newcommand{\klmcode}{$\mathcal C_{\klm}$\xspace}
\newcommand{\newcode}{$\mathcal C_{\new}$\xspace}
\newcommand{\abccode}{$\mathcal C_{\abc}$\xspace}

\newcommand{\mathlncycode}{\mathcal C_{\lncy}} 
\newcommand{\mathrepcode}{\mathcal C_{\rep}}

\newcommand{\mathconcatcode}{\mathcal C_{\concat}}

\begin{document}

\title{Avoiding coherent errors with rotated concatenated stabilizer codes} 
\author{Yingkai Ouyang}

\email[]{oyingkai@gmail.com}

\affiliation{%
 Department of Physics and Astronomy, University of Sheffield, Sheffield, UK
}%

\begin{abstract} 
Coherent errors, which arise from collective couplings, are a dominant form of noise in many realistic quantum systems, and are more damaging than oft considered stochastic errors. Here, we propose integrating stabilizer codes with constant-excitation codes by code concatenation. Namely, by concatenating an $[[n,k,d]]$ stabilizer outer code with dual-rail inner codes, we obtain a $[[2n,k,d]]$ constant-excitation code immune from coherent phase errors and also equivalent to a Pauli-rotated stabilizer code. When the stabilizer outer code is fault-tolerant, the constant-excitation code has a positive fault-tolerant threshold against stochastic errors. Setting the outer code as a four-qubit amplitude damping code yields an eight-qubit constant-excitation code that corrects a single amplitude damping error, and we analyze this code's potential as a quantum memory.  
\end{abstract}

\maketitle

\section{Introduction}
Quantum error correction (QEC) promises to unlock the full potential of quantum technologies by combating the detrimental effects of noise in quantum systems. 
The ultimate goal in QEC is to protect quantum information under realistic noise models.
However, QEC is most often studied by abstracting away the underlying physics of actual quantum systems, and assumes a simple stochastic Pauli noise model, as opposed to coherent errors which are much more realistic.

Coherent errors are unitary operations that damage qubits collectively, and are ubiquitous in many quantum systems. Especially pertinent are coherent phase errors that occur on any quantum system that comprises of non-interacting qubits with identical energy levels. 
In such systems, coherent phase errors can result from unwanted collective interactions with stray fields \cite{hogan2012driving}, collective drift in the qubits' energy levels, and fundamental limitations on the precision in estimating the magnitude of the qubits' energy levels.
To address coherent errors, prior work either 
(1) analyzes how existing QEC codes perform under coherent errors without any mitigation of the coherent errors,
(2) uses active quantum control which incurs additional resource overheads to mitigate coherent errors offers partial immunity against coherent errors \cite{debroy2018stabilizer}
or (3) completely avoids coherent errors using appropriate decoherence-free subspaces (DFS) \cite{plenio1997quantum,ZaR97,LBW99,alber2001stabilizing,alber2003detected,choi2006method,jimbo2011quantum,lin2014extremal,ouyang2019permutation}.
In this paper, we focus on a family of QEC codes that are compatible with approach (3), and discuss performing QEC protocols with respect to this family of QEC codes.

To completely avoid coherent phase errors, quantum information can be encoded into a constant-excitation (CE) subspace \cite{ZaR97,alber2003detected,ouyang2019permutation}, which is a DFS of any Hamiltonian that describes an ensemble of identical non-interacting qubits. 
Given the promise of CE QEC codes to completely avoid coherent phase errors, these codes have been studied within both qubit \cite{plenio1997quantum,ZaR97,LBW99,alber2001stabilizing,alber2003detected,jimbo2011quantum,lin2014extremal} and bosonic \cite{CLY97,WaB07,BvL16,ouyang2019permutation} settings.
Such codes either additionally avoid other types of coherent errors \cite{ZaR97,LBW99}, or can combat against other forms of errors \cite{plenio1997quantum,CLY97,alber2001stabilizing,alber2003detected,WaB07,jimbo2011quantum,lin2014extremal,BvL16,ouyang2019permutation}. However, qubit CE QEC codes lack a full-fledged QEC analysis, where explicit encoding, decoding circuits and QEC circuits remain to be constructed. This impedes the adoption of CE codes in a fault-tolerant QEC setting.

In this paper, we give an accessible procedure to construct QEC codes that not only completely avoid coherent phase errors, but also support fault-tolerant quantum computation. 
Namely, we concatenate stabilizer codes $\mathcal C_{ \stab}$ with a length two repetition code $\mathcal C_\rep$, and apply a bit-flip on half of the qubits. 
We can also naturally interpret these codes within the codeword stabilized (CWS) framework \cite{CSSZ08,SSSZ09}, thereby extending the utility of CWS codes beyond a purely theoretical setting.  

Amplitude damping (AD) errors model energy relaxation, and accurately describe errors in many physical systems. 
By concatenating the four-qubit AD code \cite{LNCY97} with the dual-rail code \cite{KLM01}, we construct an eight-qubit CE code that corrects a single AD error. 
We provide this code's QEC circuits (see Fig 3 and Fig 4), and analyze its potential as a quantum memory under the AD noise model (see Fig 5).

Our work paves the way towards integrating CE codes with mainstream QEC codes. By doubling the number of qubits required, we make {\em any} quantum code immune against coherent phase errors. When coherent phase errors are a dominant source of errors, we expect CE codes to significantly reduce fault-tolerant overheads.

\begin{figure*}[htp]
\centering 
\includegraphics[width=0.95\textwidth]{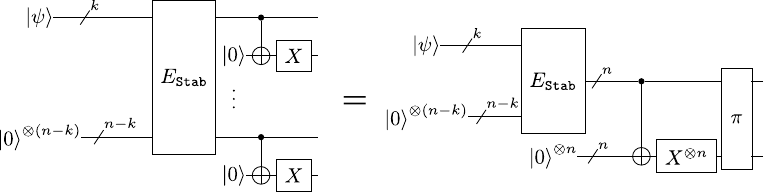}  
\caption{
Encodings of $\cC_{\stab,\klm}$ from the encoding $E_\stab$ of $\cC_{\stab}$. On the right side, CNOTs apply transversally to each pair of control and target qubits in the code blocks. The permutation $\pi$ maps the $j$th qubit in the first block of $n$ qubits to the $(2j-1)$th qubit and the $j$th qubit in the second block of $n$ qubits to the $(2j)$th qubit.
} 
\label{fig:encoding}
\end{figure*} 

\section{Results}

\subsection{Hybridizing stabilizer and CE codes}

Coherent phase errors can arise from the collective interaction of identical qubits with a classical field. Since the collective Hamiltonian of non-interacting identical qubits is proportional to $S^z = Z_1 +\dots + Z_N$ where $Z_j$ flips the $j$th qubit's phase, we model coherent phase errors with unitaries of the form $U_\theta = \exp(-i \theta S^z )$. Here, $\theta$ depends on both the interacting field's magnitude and the qubits' energy levels.

Using any CE code, we can completely avoid coherent phase errors.
This is because such codes must lie within an eigenspace of $S^z$ which is spanned by the computational basis states $|\bx \> = |x_1\> \otimes \dots \otimes |x_N\>$ for which the excitation number, given by the Hamming weight $\wt(\bx)=x_1+\dots + x_N$ of $\bx$, is constant.  
 The simplest CE code is the dual-rail code \cite{KLM01}, $\mathcal C_\klm$, with logical codewords $|0_\klm\> = |01\>$ and $|1_\klm\> = |10\>$. 
   
However $\cC_\klm$ cannot correct any errors. Therefore, we concatenate it with an $[[n,k,d]]$ stabilizer code $\cC_\stab$ to obtain a code $\cC$ with encoding circuit given in Fig.~1. Then $\cC$ is an $[[2n,k,d]]$ QEC code that is also impervious to coherent phase errors.
Now, concatenating any state 
$\smash{\sum_{\bx \in \{0,1\}^n} a_{\bx} |\bx\> \in \cC_\stab}$ with $\mathcal C_\klm$ yields 
$\smash{\sum_{\bx \in \{0,1\}^n} a_{\bx} |\varphi(\bx)\>},$
where 
$\varphi( (x_1,x_2,\dots,x_{n-1}, x_n)  ) 
= 
(x_1, 1-x_1, x_2, 1-x_2,\dots,x_{n-1},1-x_{n-1}, x_n, 1-x_n).$ 
Since $\wt(\varphi(\bx))=n$ for every $\bx \in \{0,1\}^n$, it follows that the concatenated state must be an eigenstate of $S^z$ with the same eigenvalue. 
Hence, $\cC_{\stab,\klm}$ is a CE code, and therefore avoids coherent phase errors.

The code $\cC_{\stab,\klm}$ is very similar to $\cC_{\stab,\rep}$, which is $\cC_\stab$ concatenated with a length two repetition code $\cC_\rep$ that maps $|0\>$ to $|00\>$ and $|1\>$ to $|11\>$.
Since $\cC_{\stab,\klm} = \paulirot \cC_{\stab,\rep}$ where $\paulirot= (I\otimes X)^{\otimes n}$, and $I$ and $X$ denote the identity and bit-flip operations on a qubit respectively, $\cC_{\stab,\klm}$ is equivalent to $\cC_{\stab,\rep}$ up to the Pauli rotation $\paulirot$ and we call $\cC_{\stab,\klm}$ a rotated-stabilizer code.

We can also cast $\cC_{\stab,\klm}$ within the CWS framework by deriving its word stabilizer and word operators.
Since $\cC_{\stab,\klm}$ and $\cC_{\stab,\rep}$ are equivalent up to $\paulirot$, it suffices to derive $\cC_{\stab,\rep}$'s word stabilizer and word operators.
Namely, $\cC_{\stab,\klm}$ and $\cC_{\stab,\rep}$ have identical word stabilizers generated by the stabilizer and logical $Z$ operators of $\cC_{\stab,\rep}$.
Moreover, the word operators $w_1,\dots, w_{2^k}$ of $\cC_{\stab,\rep}$ are its logical $X$ operators and the word operators $\cC_{\stab,\klm}$ are $\paulirot w_1,\dots, \paulirot w_{2^k}$. We supply explicit constructs of the word stabilizer and word operators of $\cC_{\stab,\klm}$ in ``Methods''.

\begin{figure}[htp]
\centering  
\subfigure[$L_\rep(U)$]{
\begin{minipage}[t]{0.4\linewidth}
\centering
\includegraphics[height=1.3cm]{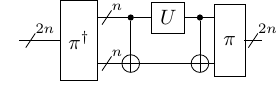} 
\end{minipage}%
}%
\subfigure[$L_\rep(U_m)$]{
\begin{minipage}[t]{0.6\linewidth}
\centering
\includegraphics[height=1.3cm]{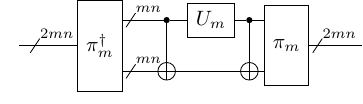} 
\end{minipage}%
}%

\caption{
Given single-qubit and multi-qubit logical operators of $\cC_\stab$ denoted by $U$ and $U_m$ respectively, we obtain corresponding logical operators for $\cC_{\stab,\rep}$ in (a) and (b) respectively. The permutation $\pi_m$ maps the $j$th qubit in the first block of $mn$ qubits to the $(2j-1)$th qubit and the $j$th qubit in the second block of $mn$ qubits to the $(2j)$th qubit.
} 
\label{fig:Lrep}

\end{figure} 
 
The code $\cC_{\stab,\klm}$ inherits its logical operators from the logical operators of $\cC_{\stab,\rep}$. 
Given any single-qubit logical operator $U$ on $\cC_\stab$, 
the corresponding unitary $\logical_\rep(U)$ on $\cC_{\stab,\rep}$ is given in Fig 2(a). Then the corresponding logical operator on $\cC_{\stab,\klm}$ is $\tilde U = \paulirot \logical_\rep(U) \paulirot$.
Similarly, given an $m$-qubit logical operator $U_m$ on $\cC_\stab$, the corresponding logical operator on $\cC_{\stab,\rep}$ is $\logical_\rep(U_m)$ (Fig~2(b)), and the corresponding logical operator on $\cC_{\stab,\klm}$ is $\paulirot^{\otimes m} \logical_\rep(U_m) \paulirot^{\otimes m}$. 
If $U$ is a tensor product of single-qubit Pauli gates, then $\tilde U$ is also a tensor product of single-qubit Pauli gates. Hence, if $\cC_\stab$ has transversal gates comprising of single-qubit Paulis, then $\cC_{\stab,\klm}$ also has corresponding transversal gates of the same form.
If $U_m$ is a diagonal unitary in the computational basis, then $\tilde U_m  = \pi_m^\dagger (U_m\otimes I^{\otimes nm} )\pi_m$ is also the logical operator on $\cC_{\stab,\klm}$.

To design error-correction procedures for $\cC_{\stab,\klm}$,
we leverage on the error-correction procedures of $\cC_{\stab,\rep}$ and the interpretation that $\cC_{\stab,\klm}$ is $\cC_{\stab,\rep}$ with an effective $R$ error.
We can extract the syndrome of a Pauli error $E$ acting on $\cC_{\stab,\klm}$ by measuring eigenvalues of Pauli observables. These Pauli observables can be generators associated with $\cC_{\stab,\rep}$'s stabilizer, and these generators 
are derived easily from the generators of $\cC_\stab$; if $G_1,\dots, G_{n-k}$ are $\cC_\stab$'s stabilizer's generators, then $\bar G_1,\dots, \bar G_{2n-k}$ generate $\cC_{\stab,\rep}$'s stabilizer, 
where $\bar G_i=\logical_\rep(G_i)$ for $i=1,\dots, n-k$ and $\bar G_{n-k+j}=Z_{2j-1} Z_{2j}$ for $j=1,\dots,n$.
We complete the QEC procedure by using measured eigenvalues of $\bar G_1,\dots,\bar G_{2n-k}$ to estimate the Pauli error $E'$ that could have occurred, and reverse its effect.

The generator $\bar G_j$'s eigenvalue on $E|\psi\>$ for $|\psi\>\in \cC_{\stab,\klm}$ when measured is $\theta_{j}= (-1)^{s_{j}}$ for some $s_{j} =0,1$. Here, $s_{j}=0$ when $\bar G_j$ and $E\paulirot$ commute and $s_{j}=1$ otherwise.
Now, denote the eigenvalue of $\bar G_j$ on $\paulirot|\psi_\klm\>\in \cC_{\stab,\rep}$ as $(-1)^{r_{j}}$ for some $r_{j} =0,1$.
Whenever $E = I^{\otimes 2n}$, we have $\smash{{\bf r}  \oplus {\bf s} = 0}$ where ${\bf r} = (r_1,\dots, r_{2n-k})$ and ${\bf s} = (s_{1}, \dots, s_{2n-k})$. 
Using $\smash{{\bf r}  \oplus {\bf s}}$, we estimate the error $E'$ that could have occurred
 on $\cC_{\stab,\klm}$.
 For this, we use any decoder $\decoder_{\stab,\rep}$ that maps a syndrome vector obtained from a corrupted state of $\cC_{\stab,\rep}$ to an estimated Pauli error.
  Such a decoder $\decoder_{\stab,\rep}$ can be a maximum likelihood decoder \cite{poulin2006optimal,pryadko2020maximum} or a belief propagation decoder \cite{leifer2008quantum,kuo2020refined,roffe2020decoding}. 
Explicitly, our code $\cC_{\stab,\klm}$'s decoder has the form 
\begin{align}
\decoder_{\stab,\klm}({\bf s}) = \decoder_{\stab,\rep}({\bf r}\oplus  {\bf s}) \label{eq:decoder-identity},
\end{align}
and thereby inherits its performance from the decoder $\decoder_{\stab,\rep}$ on the stabilizer code $\cC_{\stab,\rep}$. 

Now let us introduce some terminology related to the decoding of stabilizer codes.
Denoting the single-qubit Pauli operators as $I,X$, the phase-flip operator $Z$, and $Y =iXZ$, the set of $n$-qubit Pauli operators is $\{I,X,Y,Z\}^{\otimes n}$.
Define $\bin(P) = (\ba|\bb)$ as a $2n$-bit binary vector where $\ba = (a_1,\dots, a_n) $ and $\bb = (b_1,\dots, b_n) $ are $n$-bit binary vectors such that $P= w X^{a_1} Z^{b_1} \otimes \dots \otimes X^{a_n} Z^{b_n}$ for some $w=\pm 1, \pm i$.
Given any two Pauli matrices $P$ and $P'$ with binary representations  $\bin(P) =  ( \ba , \bb)  $ and $\bin(P') =   ( \ba' , \bb') $, their symplectic inner product \cite{CRSS98} over $\mathbb F_2$ is defined to be $\< \bin(P), \bin(P')\>_{\rm sy} = \ba \cdot \bb' +  \ba' \cdot \bb$.

To see how to decode our concatenated code, note that
\begin{align}
r_{j} &= \< \bin(\bar G_j) , \bin(\paulirot) \>_{\rm sy},\notag\\
s_{j} &= \< \bin(\bar G_j) , (\bin(E)+\bin(\paulirot) )\>_{\rm sy}, 
\end{align}
 By linearity of the inner product, it follows that 
$\smash{r_j \oplus s_j = \< \bin(\bar G_j) , \bin(E) \>_{\rm sy}}$.
This shows that $(-1)^{r_j \oplus s_j}$ is equal to the eigenvalue of $G_j$ when measured on $\paulirot |\psi\>$, the latter of which is a state in $\cC_{\stab,\rep}$, from which we can deduce \eqref{eq:decoder-identity}. 

When stochastic errors evolve under the influence of $U_\theta=\exp(-i\theta S^z)$, their weight is preserved.
First, note that 
\begin{align}
U_\theta 
&= \prod_{j=1}^N \exp(-i\theta Z_j)  = \exp(-i\theta Z)^{\otimes N}.
\end{align}
Then, for any $N$-qubit Pauli matrix $P=P_1 \otimes \dots \otimes P_N$,
we have that 
\begin{align}
\tilde P = U_\theta  P U_\theta ^\dagger
= \bigotimes_{j=1}^N  \exp(-i\theta Z) P_j  \exp(i\theta Z).
\end{align}
When $P_j=I$ or $Z$, we clearly have $\exp(-i\theta Z) P_j  \exp(i\theta Z) = P_j$.
When $P_j =X$ or $Y$, we have $\exp(-i\theta Z) P_j  \exp(i\theta Z) =\exp(-2i\theta Z)  P_j$.
For any value of $\theta$, $\exp(-2i\theta Z)  X$ and $\exp(-2i\theta Z)  Y$ are never the identity operator.
Hence we can see that the weight of $\tilde P$ is identical to the weight of $P$.
By performing stabilizer measurements, the error $\tilde P$ gets projected randomly onto some Pauli of weight equal to the weight of $P$, if this weight is no greater than half of the code's distance, it can be corrected according to the earlier-described decoding procedure.

\begin{figure*}[htp]
\centering \includegraphics[width=0.98\textwidth]{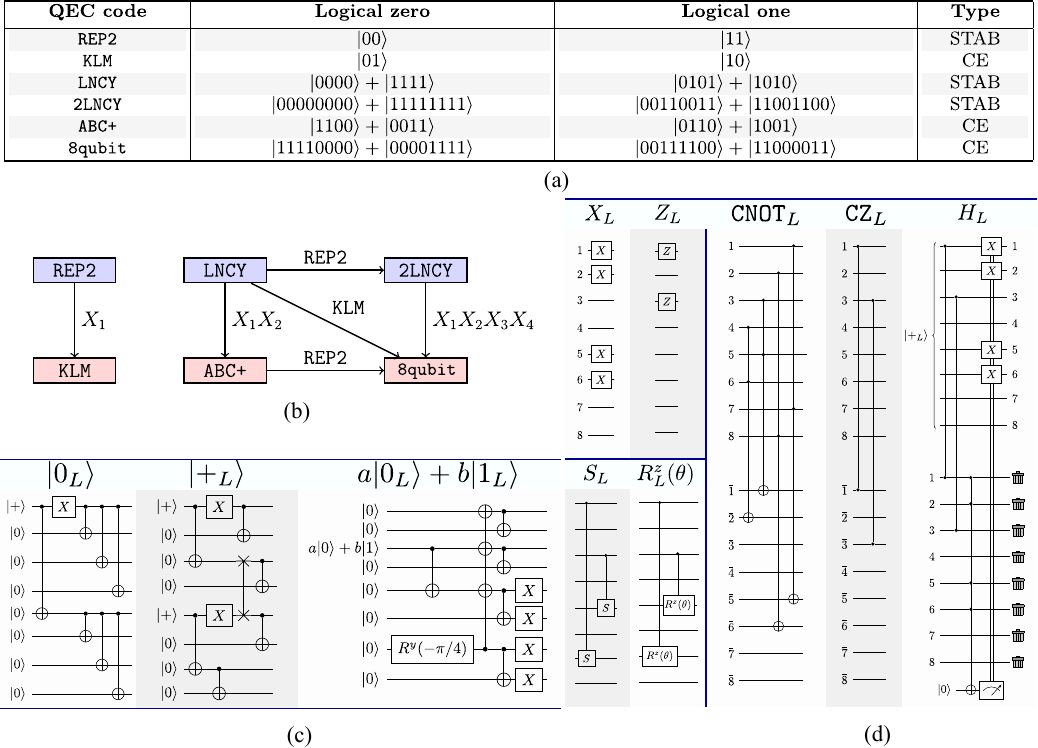}  
\caption{ (a) A table of various CE and stabilizer codes. The logical codewords are listed without their normalization factors. \rep is the two-qubit repetition code, \klm is the dual-rail code [18], \lncy code is the four-qubit AD code [17, \footnote{The four-qubit AD code is also a subcode of the [[4,2,2]] code.}] up to a permutation of qubits, \abc is a four-qubit CE code [6], \concat is \lncy concatenated with \rep and is a step to obtain our construct, and the \new code is our eight-qubit code.  
(b)  We depict the relationship between the codes in (a) pictorially. Here $X_j$ denotes a bit flip on the $j$th qubit. 
(c) State preparation circuits for \newcode, such as $|0_L\>$ and $|+_L\>$ and the logical encoding of an arbitrary logical codestate.
(d) Logical computations on \newcode are depicted. 
Here, $R^z(\theta) = e^{i Z \theta}$. The logical Hadamard is performed via logical gate-teleportation after preparing a logical $|+_L\>$ ancilla.
} \label{fig:table-encode-compute} 
\end{figure*}

We now show that $\cC_{\stab,\klm}$ has a positive fault-tolerant threshold when $\cC_\stab$ is a Calderbank-Shor-Steane (CSS) code \cite{CSS97,nielsen-chuang} that encodes a single logical qubit and has transversal logical Pauli $I,X,Y$ and $Z$ gates given by $\bar I = I ^{\otimes n}$, $\bar X = X^{\otimes n}$, $\bar Y = Y^{\otimes n}$ and $\bar Z = Z^{\otimes n}$ respectively. (also with transversal Hadamard.)
First, $\cC_{\stab,\klm}$ has transversal logical Pauli and controlled-not (CNOT) gates. Then $\cC_{\stab,\rep}$ has transversal logical $X$ and $Z$ gates given by 
$\smash{\bar X_{\rep} = \bar X^{\otimes 2}=X^{\otimes 2n}}$ and 
$\smash{\bar Z_\rep = \pi(\bar Z\otimes \bar I)\pi^\dagger}$ respectively, and logical CNOT gate $\overline{\rm CNOT}_{\rep}$ given by $2n$ transversal CNOT gates.
Thus $\cC_{\stab,\klm}$ has its logical $X$ and $Z$ operators given by 
$\smash{\bar X_{\klm}= R \bar  X_{\rep} R = X^{\otimes 2n}}$ and 
$\smash{\bar Z_{\klm}= R \bar  Z_{\rep} R = (-1)^n Z_{\rep}}$ respectively. 
Furthermore, the logical CNOT gate of $\cC_{\stab, \klm}$ has the form 
$\overline{\rm CNOT}_{\klm} = (R\otimes R) \overline{\rm CNOT}_{\rep}  (R\otimes R)  =
\overline{\rm CNOT}_{\rep}.$
Second, since we can perform these transversal CNOTs and have stabilizers that correspond to a CSS code, we can measure syndromes and logical Paulis fault-tolerantly using Steane's method for CSS codes \cite{steane1997active}. Relying on gate-teleportation techniques \cite{ZLC00}, we can implement all Clifford and non-Clifford gates fault-tolerantly.
Since the fault-tolerant logical operations will have a finite number of circuit components, using the method of counting malignant combinations in extended rectangles \cite{AGP05} yields a positive fault-tolerant threshold for stochastic noise.

 \subsection{An amplitude damping CE code}

The simplest CE code that detects AD errors is the four-qubit $\cC_\abc$ code \cite{alber2001stabilizing}. 
AD errors are introduced by an AD channel $\mathcal A_\gamma$ which has Kraus operators $A_0 = |0\>\<0| + \sqrt{1-\gamma	} |1\>\<1|$ and $A_1 = \sqrt \gamma |0\>\<1|$. These Kraus operators model the damping an excited state's amplitude and the  relaxation of an excited state to the ground state with probability $\gamma$.  
While $\cC_\abc$ detects a single AD error, it cannot correct any AD errors. Other CE codes that can correct some AD errors have been designed, but either have overly complicated encoding and QEC circuits \cite{plenio1997quantum}, or lack explicit QEC circuits \cite{LBW99,alber2003detected,choi2006method,jimbo2011quantum,lin2014extremal}.

\begin{figure*}[htp]
\centering 
\includegraphics[width=0.98\textwidth]{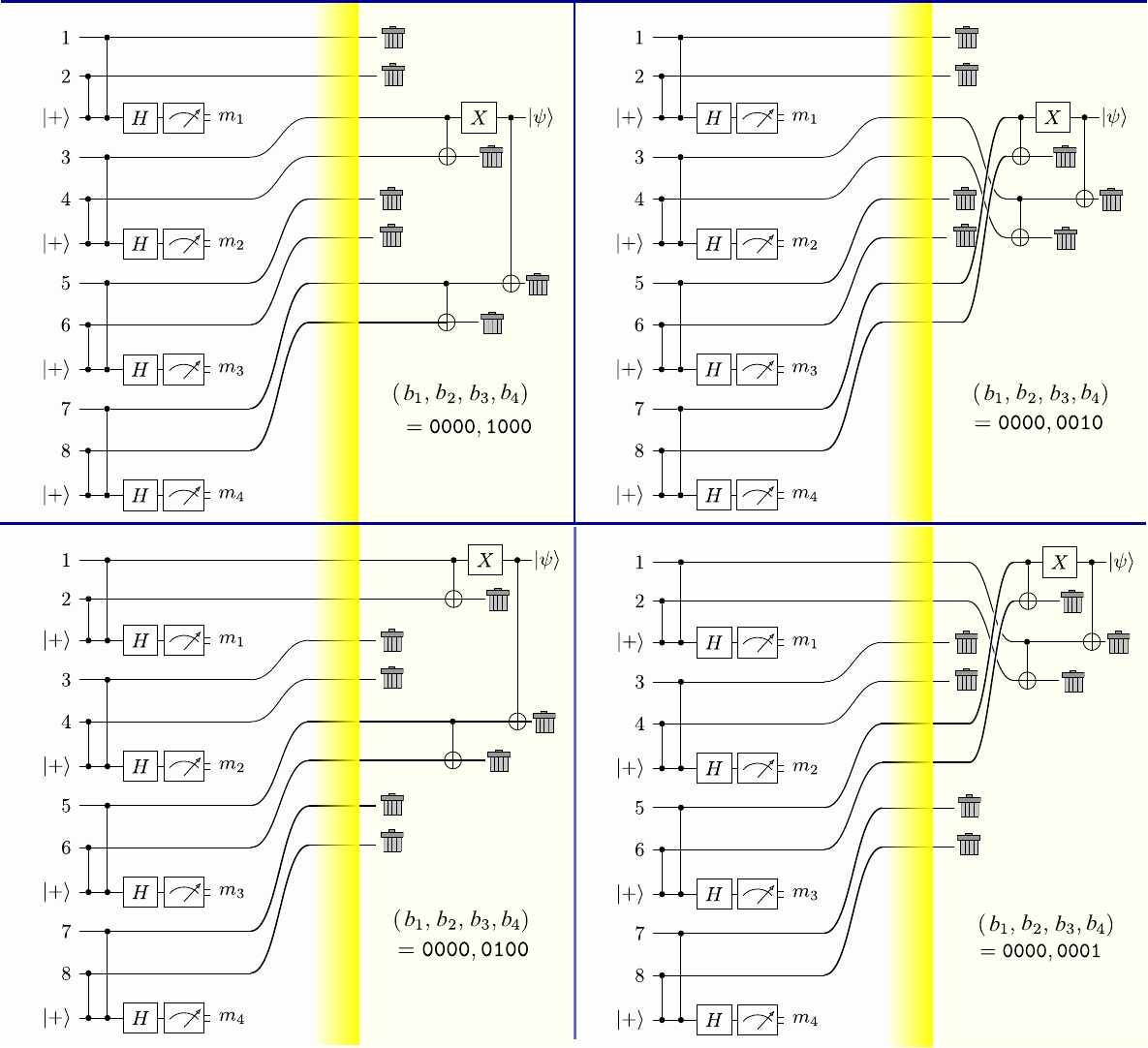}  
\caption{Syndrome extraction and decoding of \newcode. The syndrome vector is ${\bf b} =  (b_1,b_2,b_4,b_4) = (1- (m_1,m_2,m_3,m_4))/2$. If the Hamming weight of the syndrome vector is one, we can still correctly decode the logical qubit. For this, we discard four qubits and subsequently employ the same decoding circuit up to a permutation.  If the Hamming weight of ${\bf b}$ is 0, we can use any of the above decoding circuits.
} \label{fig:decode}
\end{figure*}

Here, we present a CE code that is the concatenation of the four-qubit AD code \lncycode \cite{LNCY97} with $\cC_\klm$, and permute the qubits to get \newcode with logical codewords
\begin{align}
|0_L\> &= (|11110000\> + 
|00001111\>)/\sqrt 2 \notag\\
|1_L\> &= (|00111100\> + |11000011\>)/\sqrt 2 .
\label{eq:8qubit-newcode}
\end{align}
We elucidate the connection between \lncycode, \abccode, \newcode, \klmcode and \repcode in Fig.~3(b).
We prove that \newcode corrects a single AD error by verifying that the Knill-Laflamme QEC criterion \cite{KnL97} holds with respect to the Kraus operators $K_1,\dots,K_8$ and $A_0^{\otimes 8}$ where $K_a$ denotes an $n$-qubit operator that applies $A_1$ on the $a$th qubit and $A_0$ on each of the remaining qubits.
The simplicity of \newcode allows for the direct construction of a simple error correction strategy for AD errors, without referring to the properties of \lncycode, \abccode, and \klmcode.

In Figure 3, we illustrate accessible constructs for \newcode's encoding circuits and logical computations. In Figure 4 we give decoding procedures when an AD error is detected.
We measure the eigenvalues $m_1, m_2, m_3$ and $m_4$ of the respective operators $Z_1Z_2, Z_3 Z_4, Z_5 Z_6$ and $Z_7Z_8$ to determine if any AD error has occurred.
Denoting $b_a = (1-m_a)/2$ for $a=1,\dots,4$, we have five correctible outcomes with respect to the syndrome vector ${\bf b} = (b_1, b_2, b_3, b_4)$. 
When ${\bf b} = {\bf 0}$, the codespace is damped uniformly and no AD error has occurred.
When ${\bf b}$ has a Hamming weight equal to one, each logical codeword is mapped to a unique product state, and we can ascertain that exactly one AD error must have occurred. When $b_a = 1$ and the other syndrome bits are zero, an AD error must have occurred on either the $(2a-1)$th or the $(2a)$th qubit. Since the effect of an AD error on the $(2a-1)$th and $(2a)$th qubit is identical, this makes \newcode a degenerate quantum code with respect to AD errors, and explains why \newcode has five correctible outcomes as opposed to nine if it were non-degenerate. The elegant structure of the four corrupted codespaces with a single AD error aids our construction of decoding circuits for \newcode (See details in ``Methods'').

We illustrate \newcode's performance as a quantum memory assuming perfect encoding and decoding and that AD errors only occur during the memory storage. 
We calculate probabilities $\epsilon$ and $\epsilon_{\rm base}$ of having uncorrectible AD errors occurring on \newcode and an unprotected qubit after $T$ applications of $\mathcal A_{\delta}^{\otimes 8}$ and $\mathcal A_{\delta}$ respectively.
Since the transmissivity $(1-\delta)$ of an AD channel $\mathcal A_\delta$ is multiplicative under composition, $(1-\epsilon_{\rm base}) = (1-\delta)^T$ and 
 \begin{align}
\epsilon = 1- (1-\effgam)^8 - 8 \effgam (1-\effgam)^7 \le 28 \effgam^2.
\end{align}
Whenever $28 \effgam^2 \le \effgam$, it is advantageous to use \newcode.
Hence, whenever $T \le T^\star$, where  
\begin{align}
T^\star = \frac{\log(27/28)}{\log(1-\delta)},
\end{align}
using \newcode is advantageous as compared to leaving a qubit unprotected. (See Fig 5.)


\begin{figure}[htp]
\centering  
\includegraphics[width=0.45\textwidth]{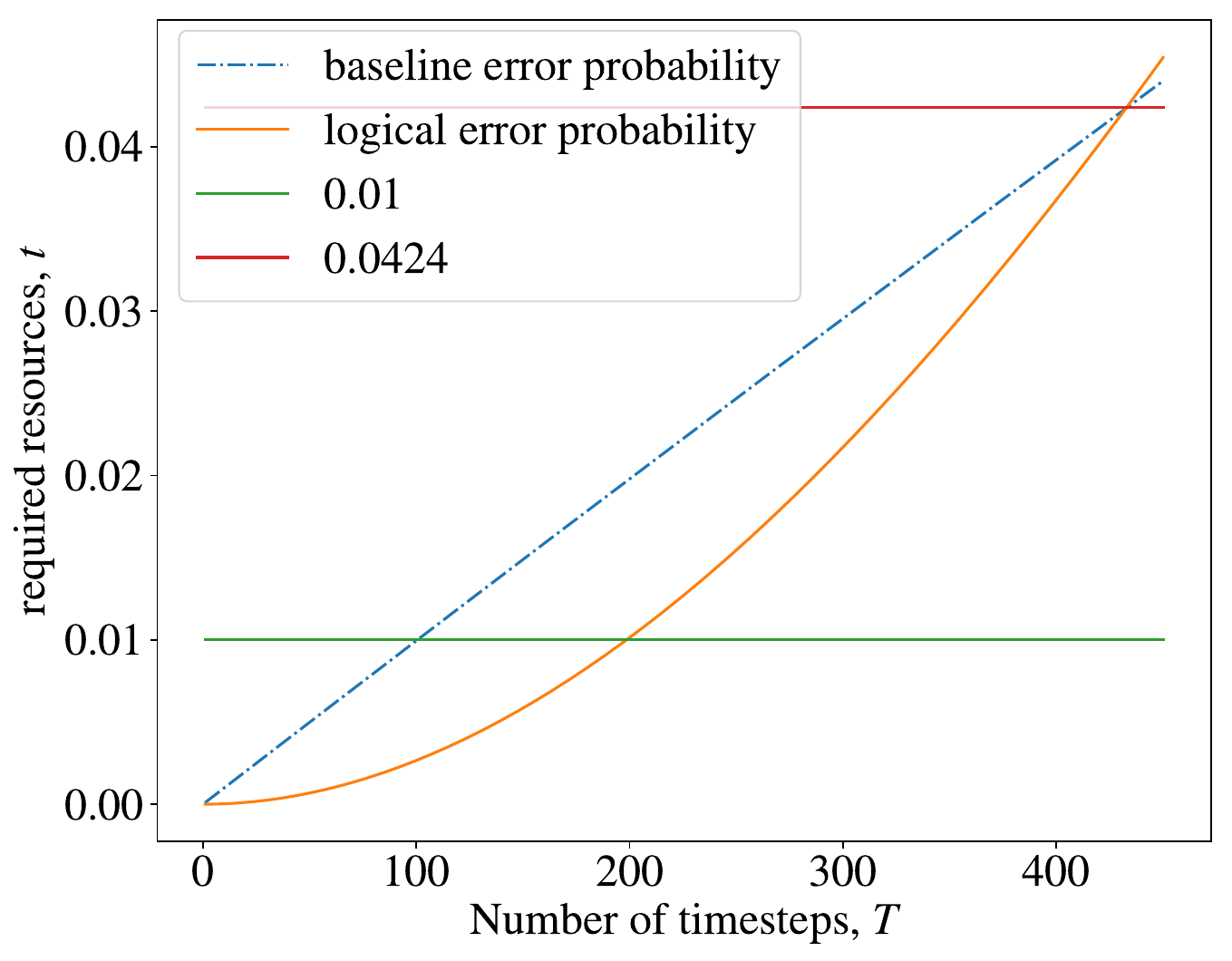} 
\caption{
Failure probability of using \newcode and an unprotected qubit versus the number of timesteps when exposed to AD errors. 
The baseline error probability is $\epsilon_{\rm base}$ and the logical error probability is $\epsilon$.
At each timestep, $\mathcal A_\delta	$ afflicts each qubit with $\delta = 10^{-4}$. When the target failure probability is 0.01, using \newcode increases the number of timesteps $T$ from about 100 to 200. When the target failure probability is over 0.0424, there is no advantage in using \newcode.
} \label{fig:bound-plots}
\end{figure}

\section{Discussion} 
 
 When coherent phase errors occur more frequently than stochastic errors, we expect CE codes to outperform generic QEC codes.
For future work, the numerical fault-tolerant thresholds of our codes can be calculated when the noise model is a convex combination of stochastic errors and coherent phase errors.
In particular, the outer codes could be chosen to be surface codes \cite{raussendorf2007fault,raussendorf2007topological,fowler2012surface}, quantum LDPC codes \cite{hypergraphcode,babar2015fifteen} and Aliferis-Preskill concatenated codes for biased noise \cite{aliferis2008fault}.  
One can also study other choices for the inner codes in our construction to obtain concatenated codes with different structures and residing in different types of decoherence-free subspaces.
For instance, we can consider other CE codes \cite{ouyang2020faster}, quantum codes that avoid exchange errors \cite{Rus00,PoR04,ouyang2014permutation,ouyang2015permutation,OUYANG201743}, and quantum codes that avoid other different errors \cite{ZaR97,lidar1998decoherence,LBW99,KLM01,choi2006method}.

\section{Methods} 
\subsection{Our CE code as a CWS code}
\label{app:CWScode}
Here, we derive the word stabilizer and word operators of our CE code 
$\cC_{\stab,\klm}$.
Now denote $S_\stab$ as the stabilizer of $\cC_\stab$ and $G_1,\dots, G_{n-k}$ as its generators.
Then the operators $\logical_\rep(G_i)  , Z_{2j-1} Z_{2j}$ generate $\cC_{\stab,\rep}$'s stabilizer where $i = 1,\dots, n-k$ and $j =1,\dots ,n$.
Denoting the logical $X$ and $Z$ operators of $\cC_\stab$ as
$\bar X_1, \dots, \bar X_k$ and $\bar Z_1, \dots, \bar Z_k$ respectively,
the logical $X$ and $Z$ operators of $\cC_{\stab,\rep}$ are given by 
$\logical_\rep(\bar X_1), \dots, \logical_\rep(\bar X_k)$
and 
$\logical_\rep(\bar Z_1), \dots, \logical_\rep(\bar Z_k)$ 
respectively.
Since $\cC_{\stab,\rep}$ is a stabilizer code, its word stabilizer of $\cC_{\stab,\rep}$ is 
\begin{align}
 W =
 \left\{
 S_{\stab,\rep}^a \prod_{j=1}^k\logical_\rep(\bar Z_j)^{z_j} : a,z_1,\dots, z_k=0,1 \right\}.
\end{align}
Since the word stabilizer of $\cC_{\stab,\klm}$ is identical to the word stabilizer of  $\cC_{\stab,\rep}$, the word stabilizer of $\cC_{\stab,\klm}$ is then given by $ W$.

Clearly, the word operators of $\mathcal C_{\stab,\rep}$ are generated by 
$\logical_\rep(\bar X_1), \dots, \logical_\rep(\bar X_k).$ 
Hence, the word operators of $\mathcal C_{\stab,\klm}$ are
\begin{align}
w_{(x_1,\dots,x_k)} =  R \prod_{j=1}^k\logical_\rep(\bar X_j)^{x_j}
\end{align}
where $x_1,\dots,x_k=0,1.$

\subsection{An amplitude damping CE code: additional details} \label{app:eight}

We now explain the connection between the codes \lncycode, \abccode, \newcode, \klmcode and \repcode as illustrated in Fig~\ref{fig:table-encode-compute}(b).
Now recall that the four-qubit amplitude damping code \cite{LNCY97} has logical codewords
\begin{align}
|0_{\lncy}\> &= (|0000\> + |1111\>)/\sqrt 2\\
|1_{\lncy}\> &= (|1100\> + |0011\>)/\sqrt 2.
\end{align}
Concatenating this with the dual-rail code $\cC_\klm$ gives the code
\begin{align}
|0_{\lncy,\klm}\> &= (|01010101\> + |10101010\>)/\sqrt 2\\
|1_{\lncy,\klm}\> &= (|10100101\> + |01011010\>)/\sqrt 2.
\end{align}
It is visually easier to work with a code if we collect the odd and even qubits in separate blocks of four qubits. 
We can achieve this by applying the permutation $\pi^\dagger$, which maps qubits 1,3,5,7 to qubits 1,2,3,4 and qubits 2,4,6,8 to qubits 5,6,7,8, to get our code with logical codewords
\begin{align}
|0_L\> &= (|00001111\> + |11110000\>)/\sqrt 2\\
|1_L\> &= (|11000011\> + |00111100\>)/\sqrt 2.
\end{align}
Note that the above code can be obtained from the four-qubit code $\cC_\abc$ with logical codewords
\begin{align}
|0_{\abc}\> &= (|0011\> + |1100\>)/\sqrt 2\\
|1_{\abc}\> &= (|1001\> + |0110	\>)/\sqrt 2,
\end{align}
after concatenation with \repcode.
Note that by concatenating \lncycode with \repcode, we get a concatenated code $\mathconcatcode = \mathlncycode \circ \mathrepcode$ with logical codewords
\begin{align}
|0_{\concat}\> &= (|0000 0000\> + |1111 1111\>)/\sqrt 2, \notag\\
|1_{\concat}\> &= (|0011 0011\> + |1100 1100\>)/\sqrt 2 .
\end{align}
Since the stabilizer code $\cC_\concat$ is equivalent to \newcode up to a Pauli rotation given by $ X^{\otimes 4} \otimes I^{\otimes 4} $, we can interpret \newcode as a rotated concatenated stabilizer code. 

To encode an arbitrary single-qubit logical state into \newcode, we concatenate the encoding circuits of \lncycode and \repcode, and apply a Pauli rotation. 
Quantum circuits can be further simplified when encode the logical stabilizer states $|0_L\>$ and $|+_L\> = (|0_L\> + |1_L\>)/\sqrt 2 $.

  To show that our QEC code spanned by $|0_L\>$ and $|1_L\>$, corrects single AD errors, it suffices to verify the Knill-Laflamme QEC conditions. In particular,
we show that for $i,j=0,1$ and $a,b=1,\dots,8$ we have $\<i_L | K_a K_b |j_L\> = \delta_{i,j}\delta_{a,b} g_a$ for some real number $g_a$.
 Now let us explain the effects of correctible AD errors on \newcode.
Recall that the correctible AD errors are given by 
$K_0 = A_0^{\otimes 8}$, 
$K_1 = A_1 \otimes A_0^{\otimes 7}$, 
$K_2 = A_0 \otimes A_1 \otimes  A_0^{\otimes 6}$, \dots ,
$K_7 = A_0^{\otimes 6} \otimes A_1 \otimes  A_0$, and
$K_8 = A_0^{\otimes 7}\otimes A_1$.
Then we can see the following.
\begin{enumerate}
\item 
$K_0|0_L\>=   (1-\gamma)^2 |0_L\>$\\
$K_0|1_L\>=   (1-\gamma)^2 |1_L\>$.
\item 
$K_1|0_L\>= \sqrt \gamma \sqrt{(1-\gamma)^3} |{01}110000\>$\\
$K_1|1_L\>=\sqrt\gamma \sqrt{(1-\gamma)^3}|{01}000011\>$.
\item 
$K_2|0_L\>=\sqrt\gamma \sqrt{(1-\gamma)^3} |{10}110000\>$\\
$K_2|1_L\>=\sqrt\gamma \sqrt{(1-\gamma)^3}|{10}000011\>$.
\item 
$K_3|0_L\>=\sqrt\gamma \sqrt{(1-\gamma)^3}|11{01}0000\>$\\
$K_3|1_L\>=\sqrt\gamma \sqrt{(1-\gamma)^3}|00{01}1100\>$.
\item 
$K_4|0_L\>=\sqrt\gamma \sqrt{(1-\gamma)^3}|11{10}0000\>$\\
$K_4|1_L\>=\sqrt\gamma \sqrt{(1-\gamma)^3}|00{10}1100\>$.
\item 
$K_5|0_L\>=\sqrt\gamma \sqrt{(1-\gamma)^3}|0000{01}11\>$\\
$K_5|1_L\>=\sqrt\gamma \sqrt{(1-\gamma)^3}|0011{01}00\>$.
\item 
$K_6|0_L\>=\sqrt\gamma \sqrt{(1-\gamma)^3}|0000{10}11\>$\\
$K_6|1_L\>=\sqrt\gamma \sqrt{(1-\gamma)^3}|0011{10}00\>$.
\item 
$K_7|0_L\>=\sqrt\gamma \sqrt{(1-\gamma)^3}|000011{01}\>$\\
$K_7|1_L\>=\sqrt\gamma \sqrt{(1-\gamma)^3}|110000{01}\>$.
\item 
$K_8|0_L\>=\sqrt\gamma \sqrt{(1-\gamma)^3}|000011{10}\>$\\
$K_8|1_L\>=\sqrt\gamma \sqrt{(1-\gamma)^3}|110000{10}\>$.
\end{enumerate}
In the above, we can see that the effect of $K_{2j-1}$ is identical to $K_{2j}$ for $j = 1,\dots,4$. Hence there are only five unique correctible outcomes that correspond to the correctible errors $K_0, K_1, K_3, K_5$ and $K_7$. 
Each of these correctible outcomes are clearly orthogonal. Hence to perform quantum error correction, it suffices to rotate the orthogonal corrupted codespaces back to the original codespace.

Now, to extract the error syndrome, it suffices to measure the stabilizers 
$Z_{2j-1},Z_{2j}$ for $j=1,2,3,4$. These stabilizer measurements leave the codespace afflicted with correctible AD errors unchanged, and measure the parity of the $(2j-1)$th and $(2j)$th qubits. We can then make the following decisions.
\begin{enumerate}
\item If the parity of the all blocks is even, then we can ascertain that no AD error has occured, which corresponds to the effect of the Kraus operator $K_0$. 
\item
If the parity of the first and second qubit is odd, while the parity of the remaining blocks is even, then we can ascertain that either $K_1$ or $K_2$ has occured.
\item 
If the parity of the third and fourth qubit is odd, while the parity of the remaining blocks is even, then we can ascertain that either $K_3$ or $K_4$ has occured.
\item
If the parity of the fifth and sixth qubit is odd, while the parity of the remaining blocks is even, then we can ascertain that either $K_5$ or $K_6$ has occured.
\item
If the parity of the seventh and eight qubit is odd, while the parity of the remaining blocks is even, then we can ascertain that either $K_7$ or $K_8$ has occured.
\end{enumerate}
The structure of the corrupted codespaces allows us to decode them into a physical qubit by first discarding four qubits, and subsequently employing the same decoding circuit up to a permutation.

\section*{Acknowledgments} 
YO acknowledges support from the EPSRC (Grant No. EP/M024261/1) and the QCDA project (Grant No. EP/R043825/1)) which has received funding from the QuantERA ERANET Cofund in Quantum Technologies implemented within the European Union’s Horizon 2020 Programme.

\bibliography{../../permutations}{}
\bibliographystyle{naturemag}

\end{document}